\documentclass{jps-cp}
\usepackage{txfonts} 
\usepackage[binary-units]{siunitx}
\usepackage{subfigure}
\usepackage{newtxtext}
\usepackage{multicol}
\usepackage{booktabs}

\title{MuPix10: First Results from the Final Design}

\author{
Heiko \textsc{Augustin}$^{1}$,
Niklaus \textsc{Berger}$^{2}$,
Sebastian \textsc{Dittmeier}$^{1}$,
David  Maximilian \textsc{Immig}$^{1}$,
Dohun \textsc{Kim}$^{1}$,
Lukas \textsc{Mandok}$^{1}$,
Annie  \textsc{Meneses Gonzalez}$^{1}$,
Marius \textsc{Menzel}$^{1}$,
Lars Olivier Sebastian \textsc{Noehte}$^{1,\dagger}$,
Ivan \textsc{Peri\'c} $^{3}$,
Alexander \textsc{Schmidt}$^{1}$,
Andr\'e \textsc{Sch\"oning}$^{1}$,
Luigi \textsc{Vigani}$^{1}$,
Alena \textsc{Weber}$^{1,3}$,
Benjamin \textsc{Weinl\"ader}$^{1}$
				}

\inst{$^{1}$Physikalisches Institut, Universit\"at Heidelberg,\\ Im Neuenheimer Feld 226, 69120 Heidelberg, Germany\\
$^{2}$Institut f\"ur Kernphysik, Johannes Gutenberg-Universit\"at Mainz,\\ Johann-Joachim-Becherweg 45, 55128 Mainz, Germany\\
$^{3}$Institut f\"ur Prozessdatenverarbeitung und Elektronik, KIT,\\ Hermann-von-Helmholtz-Platz 1, 76344 Eggenstein-Leopoldshafen, Germany\\
$^{\dagger}$ Now UZH and Paul Scherrer Institut , Switzerland
}

\email{augustin@physi.uni-heidelberg.de}

\recdate{November 24, 2020}

\abst{Many years of research and development of High Voltage Monolithic Active Pixel Sensors (HVMAPS) have culminated in the final design for the Mu3e pixel sensor.
MuPix10 is a fully monolithic sensor with an active pixel matrix size of $20\times\SI{20}{\milli\meter\squared}$ produced in the \SI{180}{\nano\meter} HV-CMOS process at TSI Semiconductors. The pixel size is $80\times\SI{80}{\micro\meter\squared}$. Hits are read out using a column-drain architecture and sent over up to four serial links with up to \SI{1.6}{\giga\bit\per\second} each.
By means of DC/DC converters and exclusive usage of on-chip biasing, MuPix10 is fully operable with a minimal set of electrical connections. This is an integral requirement by the Mu3e experiment since it enables the construction of ultra-thin pixel modules with 1\,\textperthousand\ of a radiation length per
layer. First results from lab characterisation and testbeam campaigns are presented.}

\kword{HV-CMOS, HVMAPS, particle tracking detectors, monolithic active pixel sensors}

\begin{document}
\maketitle

\section{Introduction}
The Mu3e experiment \cite{Mu3e_TDR}  is searching for the lepton flavor violating decay $ \mu\rightarrow eee$ with an unprecedented sensitivity of 1 in $10^{16}$ decays. To achieve the experimental goal, an ultra-thin pixel tracker with 1\,\textperthousand\  of a radiation length per layer is being built, which handles rates up to $10^9$ muon decays per second.

The High Voltage Monolithic Active Pixel Sensor (HVMAPS) technology \cite{Peric:2007zz} was chosen to enable the construction of this detector. The combination of drift based charge collection and integrated readout on the same chip allows to build fast pixel detectors with an excellent fill-factor. Furthermore, the small active depletion region allows thinning of the sensors to \SI{50}{\micro\meter} thickness. Using commercially available processes, the MuPix chips  \cite{Augustin:2016hzx,PIXEL:MuPix7,Augustin:2016pwd,Augustin:2017guc,Augustin:2018ppf} have been developed, culminating in the first large scale HVMAPS sensor. 

With the technological challenges solved, the next step is the construction of pixel sensor modules. The MuPix10 \cite{Hiroshima19} chip fulfils the specifications to build fully operational Mu3e tracker modules. This includes the final sensor size and a minimal chip interface. In the following, the requirements imposed by the module integration and their implementation on the MuPix10 are described. First results from testbeam measurements and the characterisation of key features are presented.


\section{Mu3e Pixel Tracker Modules}

\begin{figure}[h]
\centering
\subfigure[Schematic representation of a module, integrating four long ladders.]{\includegraphics[width=.5\textwidth]{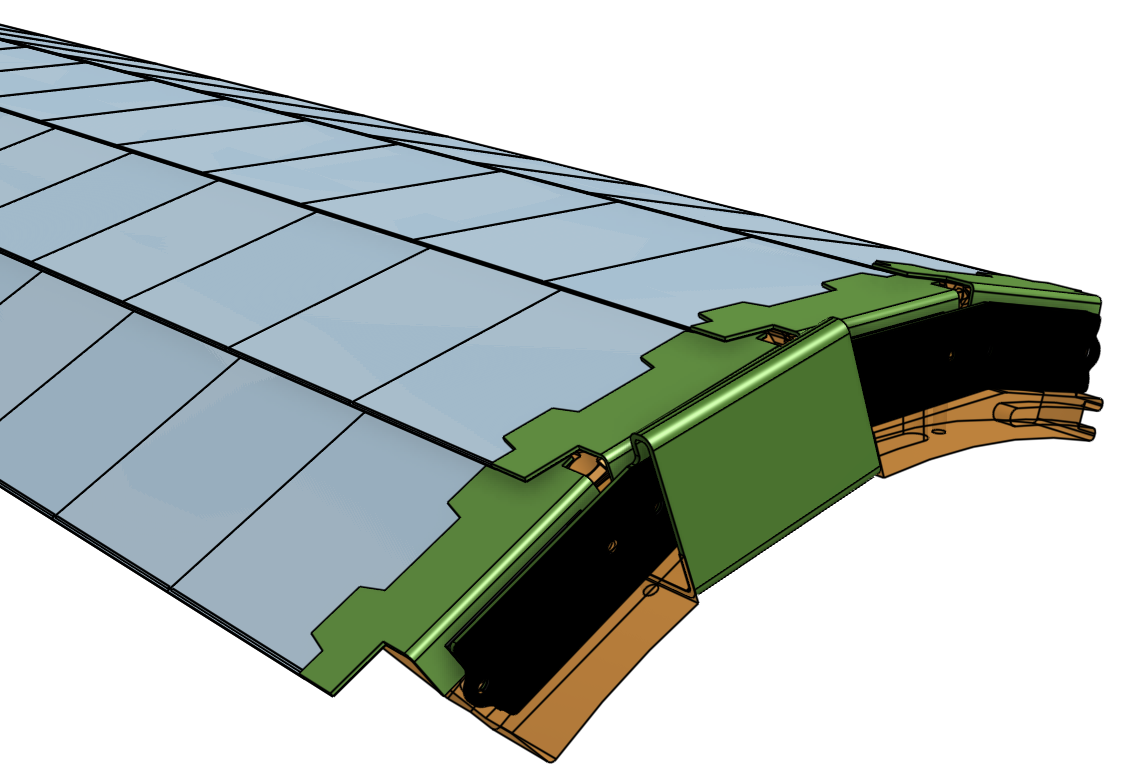}\label{fig:pixel_module}}
\qquad
\subfigure[Layer stack of the HDI. PI=polyimide, Al=aluminium.]{\includegraphics[width=.19\textwidth]{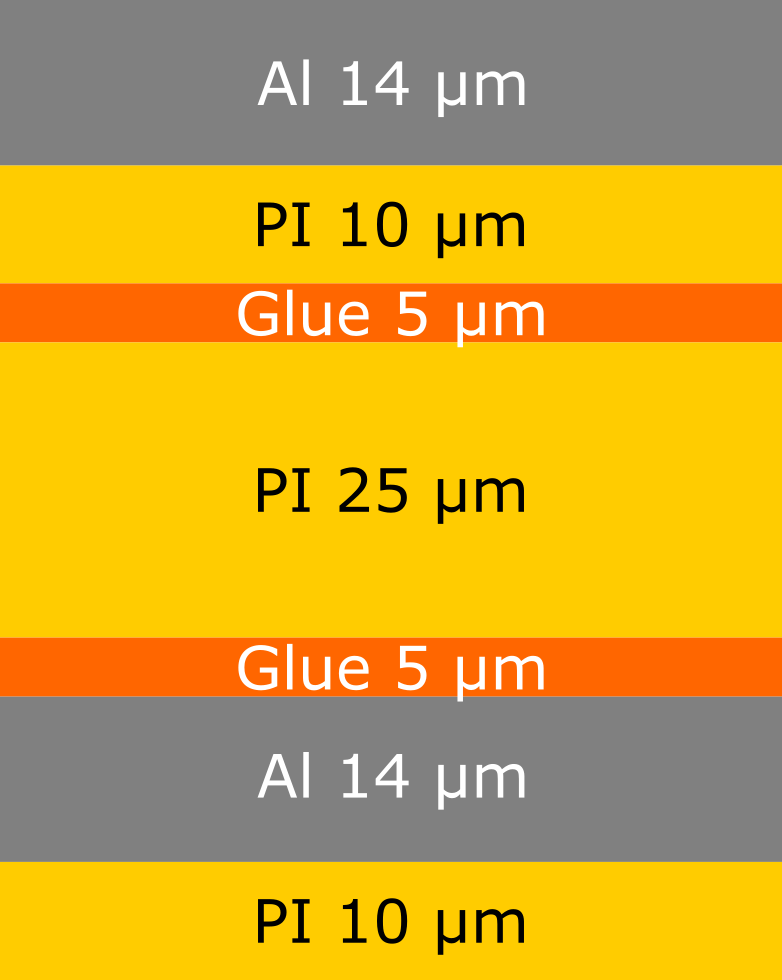}\label{fig:Stack_Al}}
\caption{CAD drawing of an Mu3e tracker module and the material stack of the HDI. \cite{Mu3e_TDR}}
\label{fig:mp10}
\end{figure}

The smallest mechanical unit of the Mu3e tracker\cite{Mu3e_TDR} is the so-called ladder. It is formed by a High Density Interconnect (HDI) circuit, which provides the electrical connections and mechanical support for 6 MuPix sensors for the inner layers of the detector and up to 18 sensors for the outer layers.
The HDIs are produced in a lithographic process by LTU Ltd. (Ukraine)\cite{LTU}, who offer a thin aluminium-polyimide technology which allows to use Single-point Tape Automated Bonding (SpTAB)\cite{SpTAB} for the sensor-HDI interconnection. The material budget of the HDI sensor stack, including adhesives, sums to 1.15\,\textperthousand\ of a radiation length. This material stack has only two aluminium layers for the HDI as depicted in figure~\ref{fig:Stack_Al}. These two layers provide power and the high-voltage bias, as well as control and data lines for the chip, which are realised as differential pairs.

The ladders are electrically divided into two halves and the MuPix sensors are read out from both ends.  Every half ladder provides two differential buses which are used for clocking and the configuration of the sensors. The sensors are read out with 9 differential data lines to both ends at a bandwidth of \SI{1.25}{\giga\bit\per\second}  per line, for the inner layers 3 per chip, in the outer layers 1 per chip. Power and ground are supplied globally, the sensors are connected in parallel. As the chips do not offer regulators for the \SI{1.8}{\volt} nominal operating voltage, voltage gradients between sensors due to ohmic losses are avoided in the design of the power lines. For the outer layers up to \SI{30}{\watt} have to be provided through this power grid, motivating a rigorous minimisation of the configuration and readout interface to allow for wider power traces.
\section{The MuPix10 chip}
In the design of MuPix10, all requirements imposed by the module framework have been implemented, leading to the first full-scale and module-ready MuPix prototype. The chip features two sets of aluminium bonding pads. One with a pad size of $200\times\SI{100}{\micro\meter\squared}$ meeting the requirements of the HDI interface for SpTAB. The second, wedge bond pads of $90\times\SI{150}{\micro\meter\squared}$ size, doubling the SpTAB pads and adding further signals, which help with the initial characterisation in  the laboratory and offer alternative ways for the chip configuration and monitoring.

MuPix10 comprises a matrix of $256\times250$ pixels with a size of $80\times\SI{80}{\micro\meter\squared}$, with a total active area of $20.48\times\SI{20.00}{\milli\meter\squared}$. The size of the die is $ 20.66\times\SI{23.14}{\milli\meter\squared}$, see figure~\ref{fig:mp10}. This includes the chip guard ring which is surrounded by an \SI{11}{\micro\meter} seal ring to minimize dead space between chips on a ladder. 

\begin{figure}[tbh]
\centering
\includegraphics[width=.4\columnwidth]{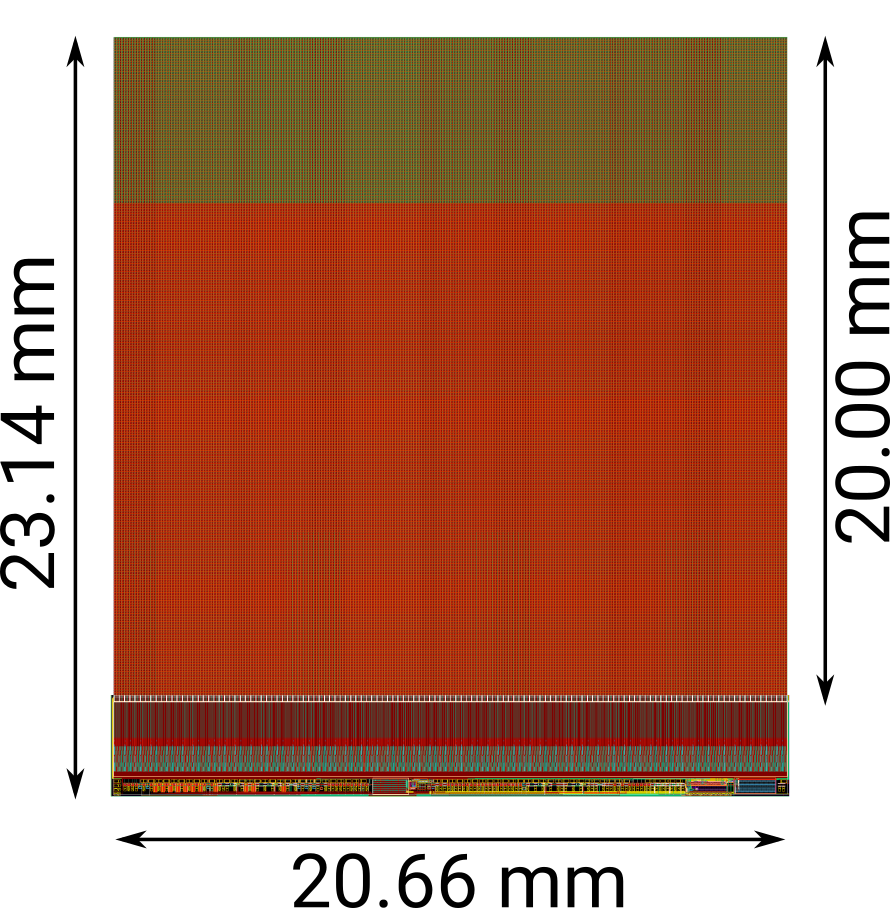}
\caption{The MuPix10 layout: Active matrix at the top and periphery at the bottom. The color change in the active matrix is a feature of the routing scheme, see section \ref{sec:crosstalk}}
\label{fig:mp10}
\end{figure}

\begin{figure}[tbh]
\centering
\includegraphics[width=\columnwidth]{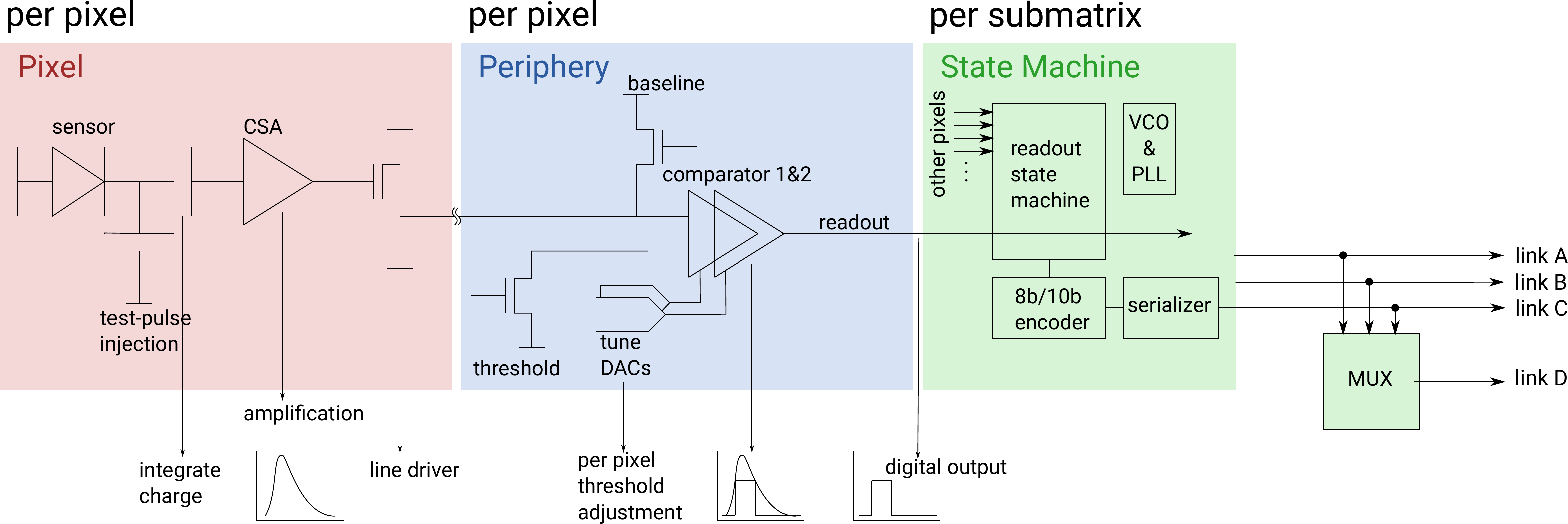}
\caption{A functional description of the MuPix architecture. From the in-pixel electronics with injection capability to the discriminator and readout infrastructure in the periphery. \cite{Mu3e_TDR}}
\label{fig:architecture}
\end{figure}

The architecture of MuPix10 follows the standard MuPix approach depicted in figure~\ref{fig:architecture}.  The deep n-well contains a folded cascode amplifier with a PMOS input transistor and a source follower as line driver, implemented as floating logic. An additional supply voltage of nominally \SI{1.2}{\volt} is required by the amplifier, which is not provided by the HDI. Therefore a voltage regulator is implemented to generate this additional voltage level from the \SI{1.8}{\volt} supply, see section \ref{sec:reg}. The pixel line driver transmits the amplified signal via a point-to-point connection to the periphery to its digital partner-cell. This signal is AC-coupled to two parallel comparators which allow to apply different threshold schemes for hit detection and time sampling. Two timestamps are stored for each hit. The rising edge is sampled with 11 bit in \SI{8}{\nano\second} bins (TS1). A second, 5~bit, timestamp (TS2) is stored on the crossing of the falling edge which allows to measure a Time-over-Threshold (ToT) by calculating the difference of the two timestamps. The sampling speed is adjustable to allow for a variable dynamic range. To ensure a correct measurement of the ToT and to maintain the level of chronology, a delay circuit is implemented which inhibits the readout of the hit after first registration for a constant time, see section \ref{sec:delay}.
The chip is split into 3 sub-matrices with 84, 86 and 86 columns, respectively, which are read out in a column-drain fashion. The serialised and 8b10b encoded hit data is sent out over a differential link with a nominal bandwidth of \SI{1.25}{\giga\bit\per\second}. There is one link per sub-matrix and an additional fourth multiplexed link which allows to send out data of all sub-matrices in a combined stream. Using the individual links a, theoretically, total rate of \SI{90}{\mega Hit\per\second} can be transmitted.

All control voltages are generated on chip, including the discriminator thresholds, making the chip operable with a minimal set of connections. To allow monitoring from outside, a programmable ADC can measure key voltages on the chip and send out the measured values via the data link to provide crucial information e.g. the threshold levels and the chip temperature.

\section{First Results}
In the following, preliminary results from the MuPix10 commissioning and tests of key features are presented.

\begin{figure}[h]
\centering
\subfigure[A MuPix10 chip mounted on a test PCB.]{\includegraphics[width=.3\textwidth]{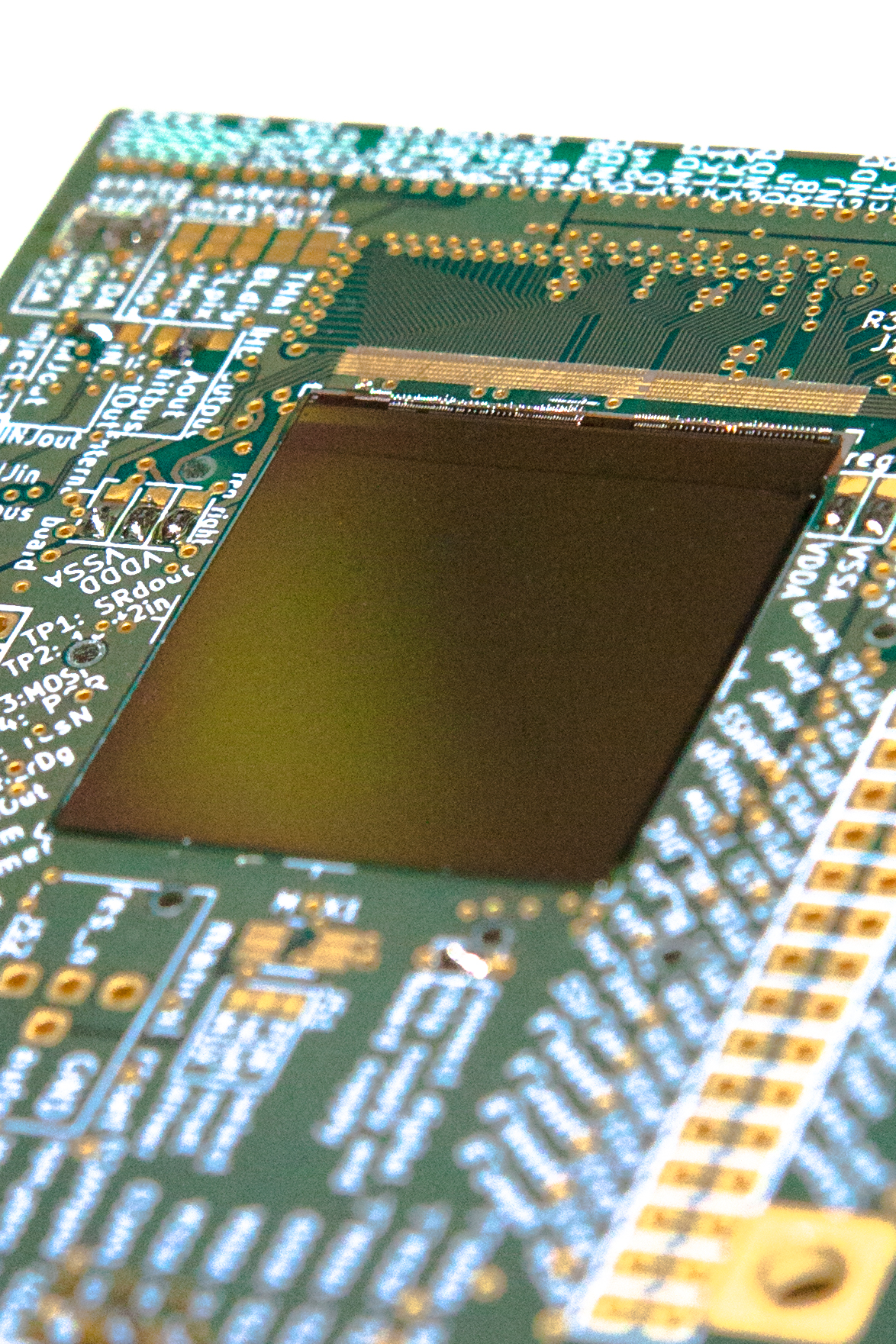}\label{fig:mp10_insert}}
\qquad
\subfigure[MuPix10 hit map observed with the DESY electron beam.]{\includegraphics[width=.6\textwidth]{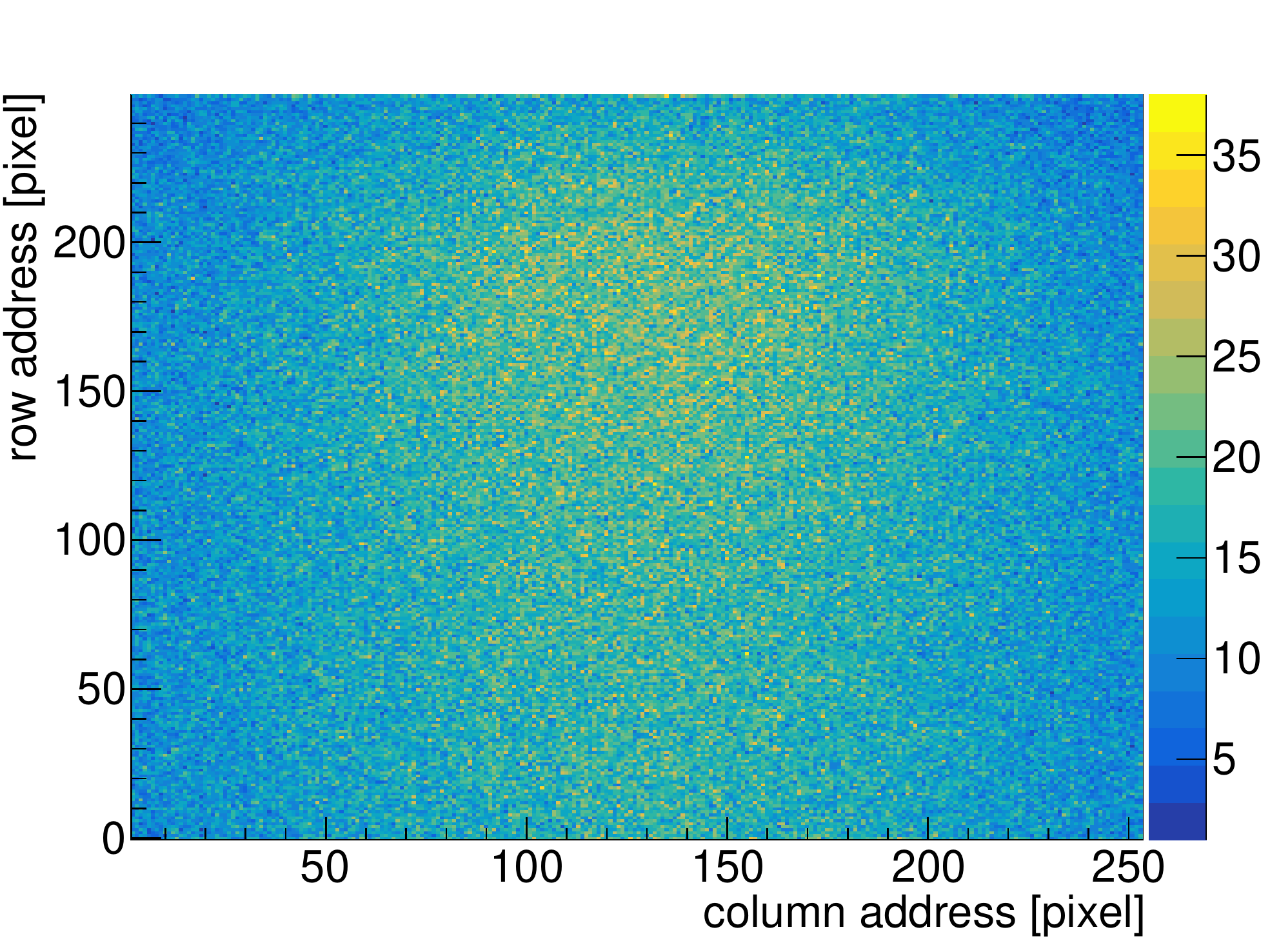}\label{fig:mp10_hitmap}}
\caption{MuPix10.}
\label{fig:mp10_hitmap}
\end{figure}

\subsection{Commissioning}
MuPix10 has been successfully commissioned in the laboratory. It is fully configurable and works reliable with the nominal reference clock of \SI{125}{\mega\hertz}, providing stable differential data links at \SI{1.25}{\giga\bit\per\second}.  At full readout speed problems with the data integrity have been observed. Therefore, it is currently throttled by a factor of 2. 
The breakdown of the deep n-well diode was measured to be around \SI{-100}{\volt} which exceeds the minimum requirement of \SI{-60}{\volt} . For a \SI{200}{\ohm\centi\meter} substrate this provides a depletion zone of $\approx\SI{40}{\micro\meter}$, which means there is no inactive bulk material for \SI{50}{\micro\meter} thin chips.
For testbeam use, a preliminary optimisation of the chip settings was performed, based on knowledge from previous chip generations and qualitative investigations with radioactive sources ($^{90}\text{Sr}$ and $^{55}\text{Fe}$). With these settings and all supply voltages provided externally, the power consumption of the chip is around \SI{190}{\milli\watt\per\centi\meter\squared}, normalised to the active area. Figure~\ref{fig:mp10_hitmap} shows a MuPix10 hitmap obtained with a \SI{3}{\giga\electronvolt} electron beam at DESY. With the large area, the core of the DESY beam is fully confined in the matrix.
Based on the obtained data, the efficiency and time resolution is currently being studied.
%

\subsection{VSSA Regulator}
\label{sec:reg}
\begin{figure}[h]
\centering
\subfigure[Functional sketch of the linear regulator.]{\includegraphics[width=.34\textwidth]{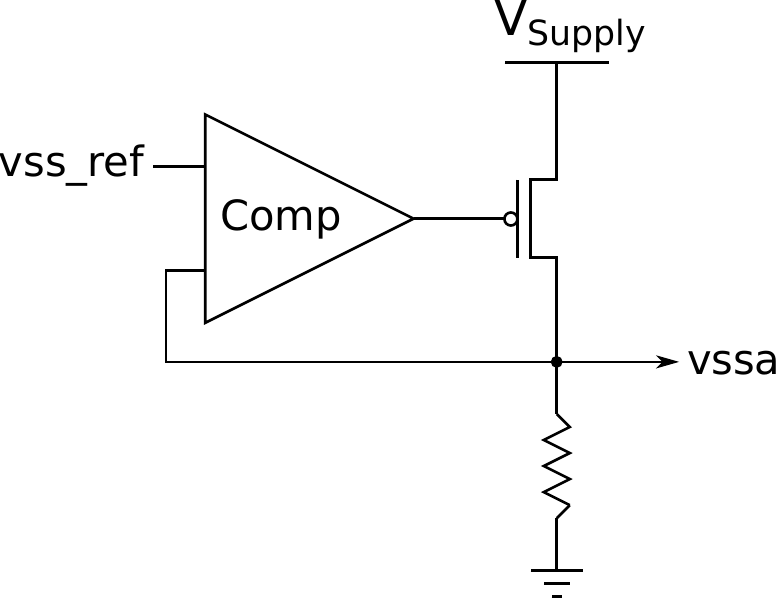}\label{fig:regulator}}
\qquad
\subfigure[VSSA voltage and supply current plotted for a scan of the regulator reference voltage (vss\_ref).]{\includegraphics[width=.55\textwidth]{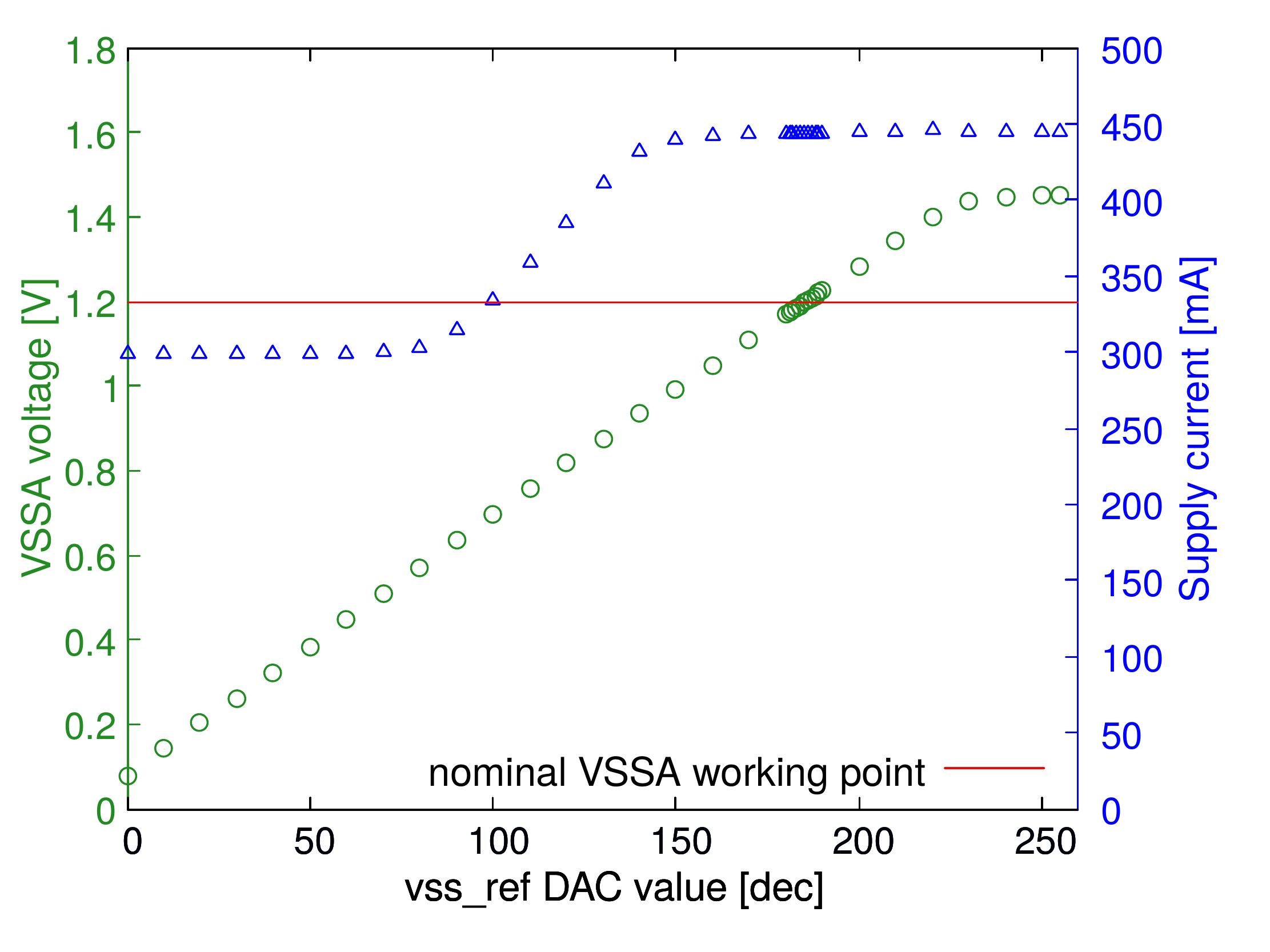}\label{fig:reg_result}}
\caption{The VSSA regulator.}
\label{fig:sub-pixel}
\end{figure}

The architecture of the regulator is sketched in figure~\ref{fig:regulator}. It is a linear series regulator which uses a differential amplifier to adjust the VSSA voltage level to a configurable reference value (vss\_ref). As the regulator acts as a self adjusting voltage divider the current drawn by the \SI{1.2}{\volt} network will dissipate additional heat in the regulating transistor leading to a power penalty.
So far no detailed study of the regulator has been performed, however, an ad-hoc attempt to run the chip with a single supply voltage, was performed successfully. Figure~\ref{fig:reg_result} shows a scan of the regulator reference voltage. The VSSA level shows a linear behaviour with a range from \SIrange{0}{1.4}{\volt} with the nominal working point of \SI{1.2}{\volt} well covered. Further, the corresponding changes in the supply current are plotted. As the amplifier is the only circuit using the VSSA voltage; the change in current represents the turn-on of the amplifier. Only for voltage levels above \SI{1}{\volt} the amplifier can be considered fully working. 
For the used settings, the total power consumption was measured to be \SIrange{200}{210}{\milli\watt\per\centi\meter\squared} which corresponds approximately to a 10\% power increase due to the regulator usage. With this, the Mu3e requirement of the sensor power consumption of $<\SI{350}{\milli\watt\per\centi\meter\squared}$ is very well met.

%

\subsection{Threshold Tuning}
All pixels are connected to two comparators with globally applied thresholds. Due to process variations the threshold behaviour of the comparators will vary and are a source of non-uniformities. In the MuPix10, both comparators are equipped with a 3 bit DAC which allows to set individual threshold to compensate the variations, the so-called trimming or tuning. Additionally, the pixels have a switch bit, which allows to mute pixels if they are uncontrollably noisy. This feature was tested and a tuning was performed successfully \cite{Menzel2020}.

\begin{figure}[h]
\centering
\subfigure[Linear voltage steps achieved with the 3 bit tuning DAC.]{\includegraphics[width=.45\textwidth]{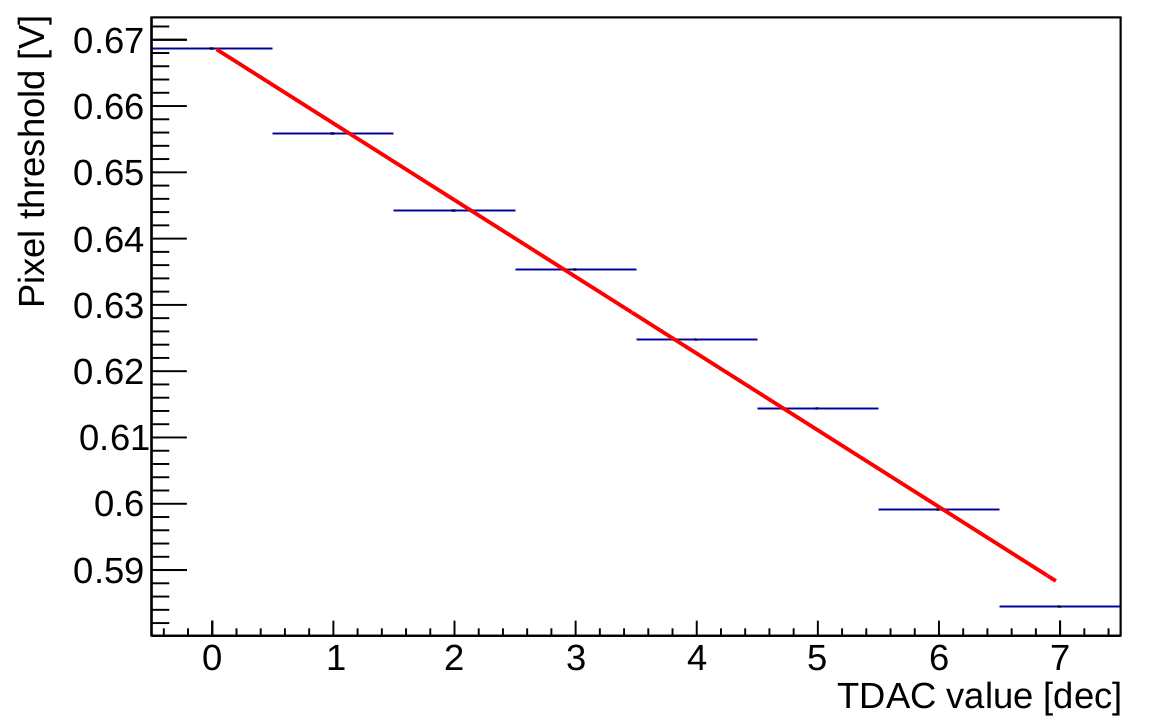}\label{fig:tuning_tdac}}
\qquad
\subfigure[The threshold dispersion before (red) and after (blue) tuning for the full MuPix10 matrix with an equivalent signal of $3000 \text{\,e}^{-}$.]{\includegraphics[width=.45\textwidth]{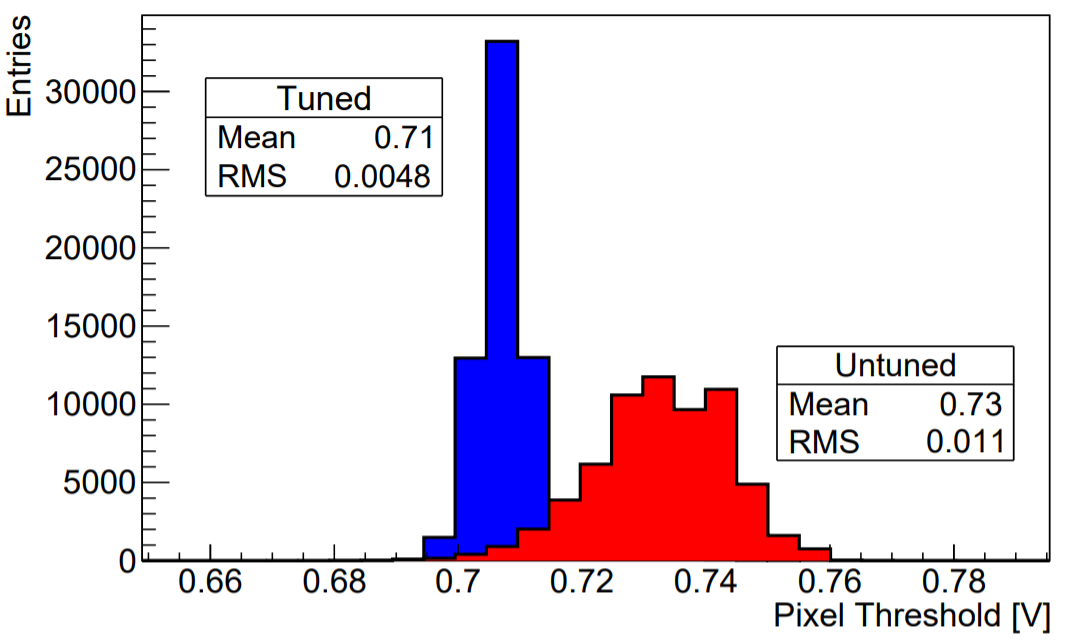}\label{fig:tuned}}
\caption{Results obtained with the threshold tuning method~\cite{Menzel2020}.}
\label{fig:sub-pixel}
\end{figure}

In this study all pixels have been stimulated with charge pulses corresponding to $3000 \text{\,e}^{-}$, using the injection infrastructure, see figure~\ref{fig:architecture}. A scan of the global threshold is performed and the amount of recognised injection pulses is measured. The resulting distribution is fitted with an s-curve function. The mean value extracted from this fit varies for different pixels. The RMS of the mean-distribution plotted in figure~\ref{fig:tuned} is a measure of the threshold dispersion. With the application of individual threshold shifts, shown in figure~\ref{fig:tuning_tdac}, the differences of the mean values can be minimised, reducing the dispersion over all pixels.

%

With this method the threshold dispersion was reduced from \SI{11}{\milli\volt} to \SI{4.8}{\milli\volt} or in electron equivalent from $240 \text{\,e}^{-}$ to $75 \text{\,e}^{-}$, see figure~\ref{fig:tuned}.
The effect of the tuning on the efficiency still needs to be investigated in testbeam campaigns. 

\subsection{ToT-sampling} 
\label{sec:delay}
As shown with the MuPix8 chip \cite{Hammerich2018,Pixel18} the time resolution of the MuPix sensors can be significantly improved by correcting timewalk by using the ToT information.
Although not necessarily required in Mu3e, the precise time information can help to ease tracking in the offline analysis. To provide correctly measured ToT values, while maintaining the data chronology and readout speed, the introduction of a new circuit is necessary.

\begin{figure}[h]
\centering
\subfigure[Inhibiting the pixel readout for a constant time, after hit recognition.]{\includegraphics[width=.345\textwidth]{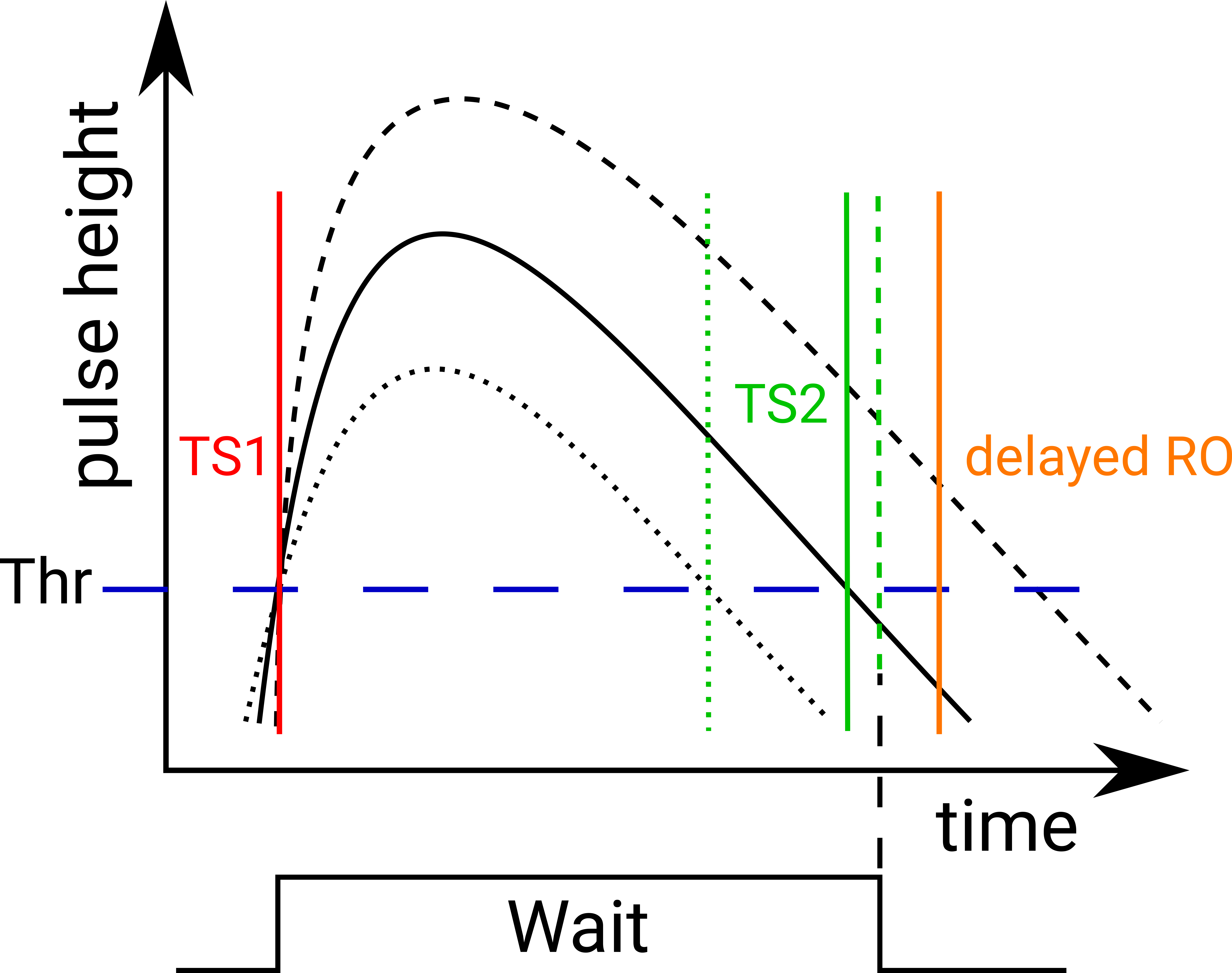}\label{fig:delay_wait}}
\qquad
\subfigure[The delay circuit logic. (The reset circuit is not shown.)]{\includegraphics[width=.38\textwidth]{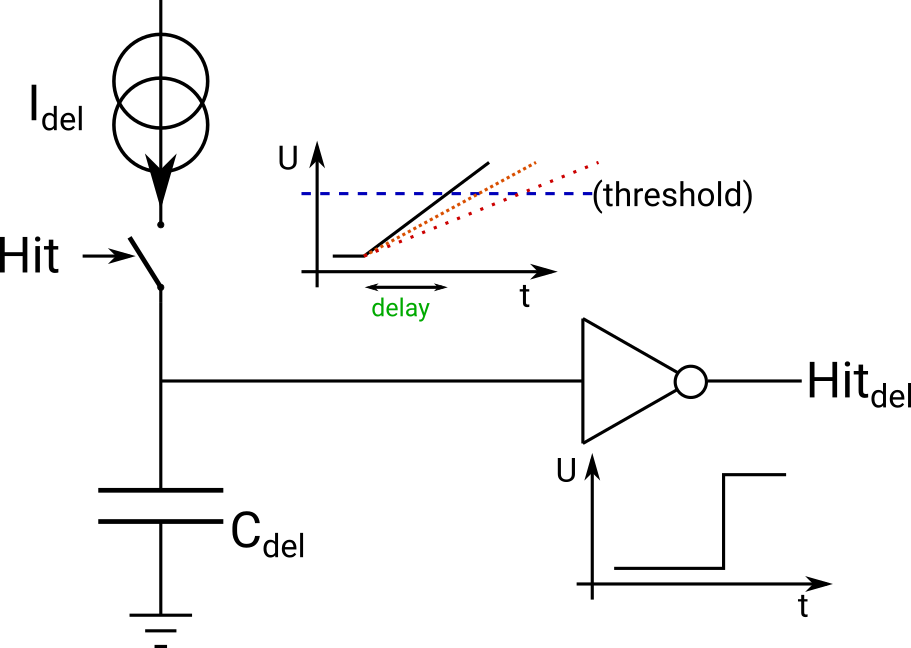}\label{fig:delay_circuit}}
\caption{Readout (RO) and timestamp sampling for different pulse lengths using a constant delay.}
\label{fig:tot-sampling}
\end{figure}

The solution chosen for MuPix10 is shown in figure~\ref{fig:delay_wait}. After the hit is registered, the readout of the hit is inhibited for a constant time. The delay is created by a new analogue delay circuit depicted in figure~\ref{fig:delay_circuit}. The registered hit enables a current source which charges up a capacitor. The increasing voltage of the capacitor is measured by a discriminating element which switches when its threshold is crossed. This output enables the readout of the pixel cell. The strength of the current source is adjustable and thus the time delay. 

\begin{figure}[tbh]
\centering
\includegraphics[width=.65\columnwidth]{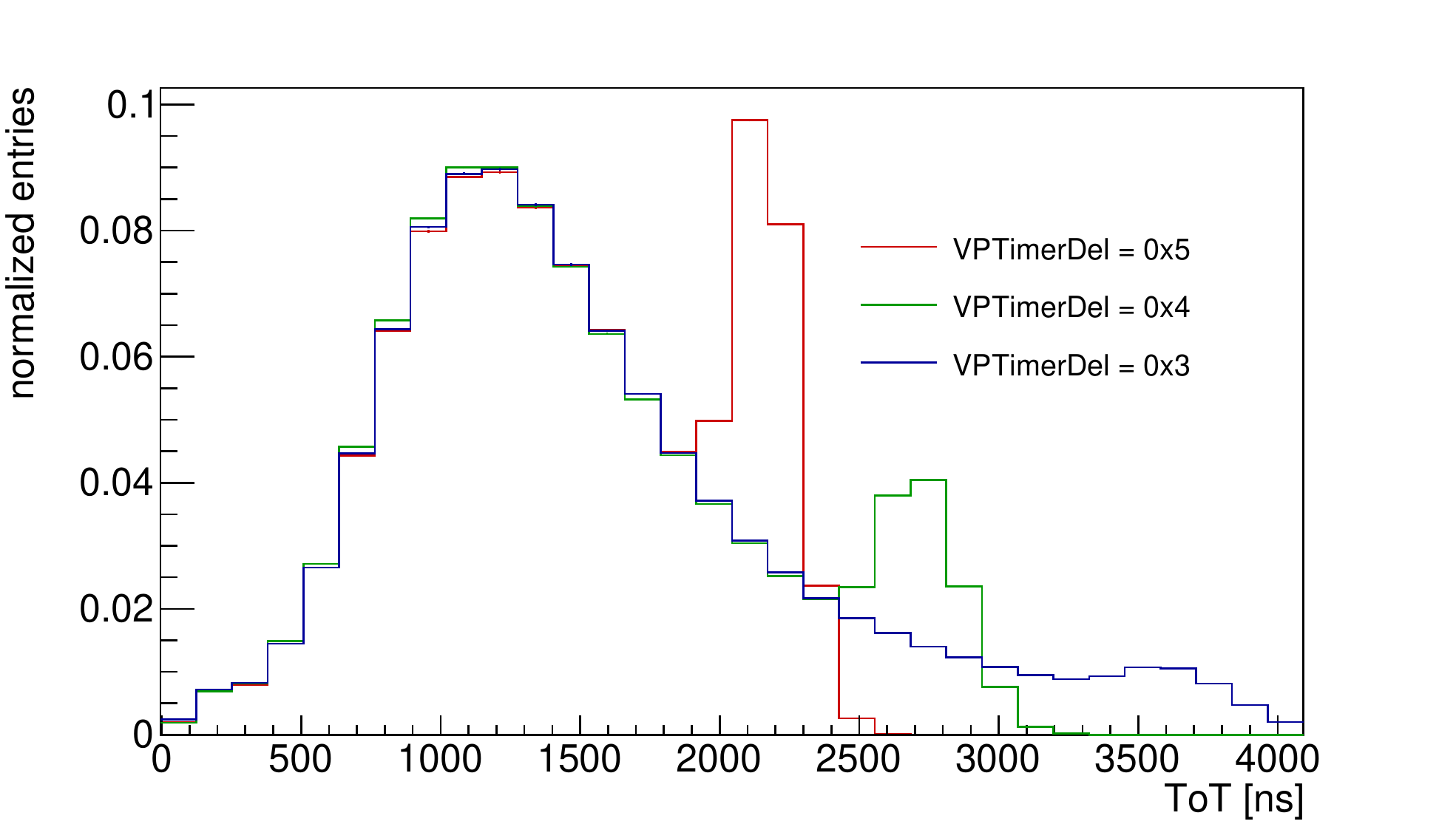}
\caption{ToT spectra obtained for 3 different delay times controlled by the VPTimerDel DAC value.}
\label{fig:delay_tot}
\end{figure}

Figure~\ref{fig:delay_tot} shows three measured ToT distributions with different chosen delays. The peak in the end of the spectrum corresponds to the delay time. If the delay time lapsed, but the ToT of the pulse would be even longer, the expiring time of the delay is sampled instead. All pulses longer than the delay are contained in this peak. The dispersion of the peak is caused by variations of the delay time over the chip and is on an acceptable level for the usage in Mu3e. As long pulses correspond to a large signal amplitude, there is no need for a timewalk correction. Hence, the delay time can be reduced below the duration of the longest pulses without loosing precision in the correction process.

This shows that the delay circuit is working as intended and the correct ToT sampling is secured. Detailed studies on the performance, sensor variations and possible position dependences of the circuit are under way.

\subsection{Signal Line Crosstalk}
\label{sec:crosstalk}
Crosstalk in a MuPix chip was first observed with the MuPix7 prototype and could be fully attributed to the effect of signal line crosstalk. Other types of crosstalk have not been observed \cite{PIXEL:MuPix7}. 
For MuPix-like chips, all pixels within one column have a point-to-point connection to the periphery which is routed within the same column. Neighbouring lines form a parallel plate capacitor, where the capacitance between those lines scales with the length they are adjacent. Via this parasitic capacitance the two lines are coupled and a signal pulse on one line creates a small crosstalk pulse in the neighbour. If it surpasses the comparator's threshold, an additional hit is created. The larger the capacitance, the larger the crosstalk pulse. This effect was observed in the MuPix8 chip which featured increasing line lengths to more than \SI{1.6}{\centi\meter} and with this increasing capacitive coupling \cite{Huth2018}. For the longest lines of the MuPix8 the crosstalk pulse had approximately 17\% of the amplitude of the initial pulse \cite{Hammerich2018}. This gives rise to a 35\% chance of crosstalk to both neighbouring lines, creating two additional hits which artificially increases the readout load. The MuPix8 only features 200 pixels within one column, while 50 more pixels are added for the Mupix10, thus further increasing the line length and therefore the crosstalk amplitude and as a consequence the amount of additional hits.

\begin{figure}[h]
\centering
\subfigure[The MuPix10 routing pattern with exemplary highlighted coupling capacities.]{\includegraphics[width=.25\textwidth]{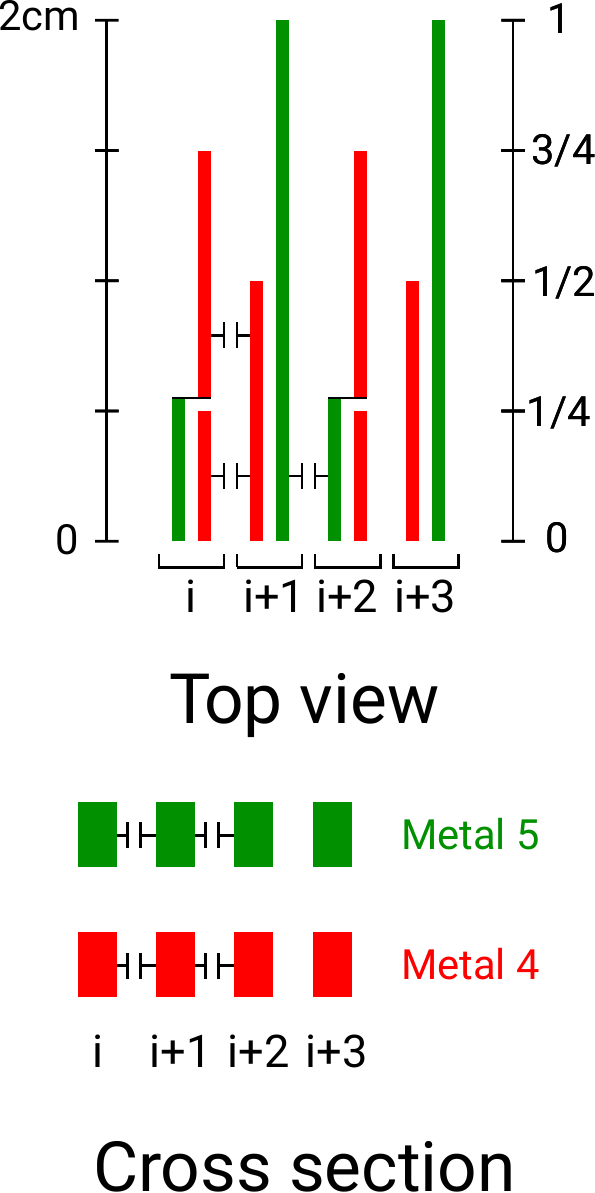}\label{fig:mp10_routing}}
\qquad
\subfigure[c][Row address self correlation. Off diagonal elements are enhanced by lowering the displayed max-value.]{\includegraphics[width=.69\textwidth]{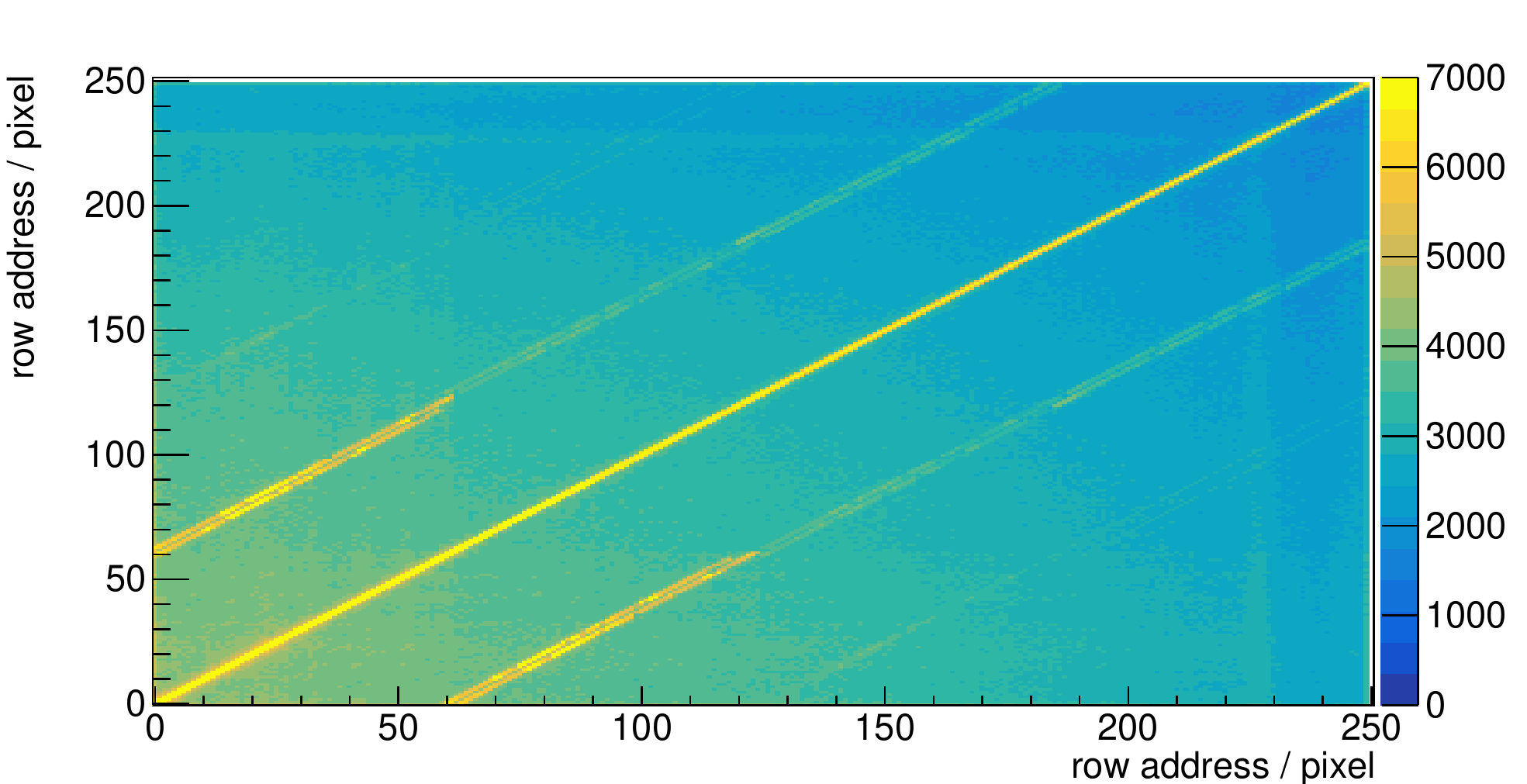}\label{fig:xtalk}}
\caption{Implementation and effect of the crosstalk routing.}
\label{fig:sub-pixel}
\end{figure}

For the MuPix10 design, two metal layers were available for pixel routing. With the usage of the full column width and 125 lines per metal layer, the smallest distance between adjacent lines is the same as in MuPix8. Two strategies were applied in the MuPix10 design to reduce crosstalk as much as possible and make it easily distinguishable from "real" sensor phenomena such as charge sharing.
For the latter, neighbouring pixels must not be routed on adjacent signal lines. By choosing an easily recognisable pattern, crosstalk can be identified and removed from the data. In case of one-sided crosstalk, the ToT information helps to identify the crosstalk hit, as the crosstalk pulses will be very short, while the initial signal is large.
The second strategy aims to reduce crosstalk by minimising the length two different signal lines are routed in close proximity. With the two metal layers available, the scheme in figure~\ref{fig:mp10_routing} allows to limit the direct neighbour length to $1/4$ of the maximal length. Following this simple picture and the extrapolation from the MuPix8 results, the new routing scheme reduces the crosstalk amplitude to 5\% of the initial pulses amplitude. Therefore reducing the amount of additional hits by crosstalk. 

In contrast to MuPix8, the signal line length is not increasing with the row position, but four discrete lengths are chosen to achieve a more uniform behaviour. Also, as a correlation between the total line capacitance and the pixel delays is expected \cite{Hammerich2018}.
Figure~\ref{fig:xtalk} shows the row address self-correlation of within one timestamp cycle. While charge-sharing will only create a broadening of the diagonal, due to the chosen routing pattern, the off-diagonal enhancements are fully attributed to crosstalk and can be easily identified. A first estimation by counting the crosstalk induced patterns and comparing them to the number of all clusters gives a total crosstalk probability of less then 1.5\% for the MuPix10. 

%
\newpage
\section{Summary and Conclusion}
MuPix10 was commissioned successfully and works reliably. As it shows no obvious performance degradation considering efficiency and time resolution it is already used as reference layers in the MuPix-Telescope \cite{Huth2018}. Key features, such as the new routing scheme, have been tested and perform as expected. It was shown that the chip can be powered with a single \SI{1.8}{\volt} supply voltage, with \SI{1.2}{\volt} generated by an on-chip regulator. This is an important milestone for the module readiness of the Mu3e pixel sensor. The next step is the production of a multi chip demonstrator.

Detailed studies of the presented features, the sensor efficiency and the time resolution are ongoing and will provide important input for the design of the MuPix11, which is going to be used for the module production. The design of MuPix11 will be very close to MuPix10 and only incorporates necessary fixes and possible performance improvements which are well reviewed and unlikely to cause a failure of the chip as a whole. 

%
%
%

\section*{Acknowledgements}
H.~Augustin, A.~Meneses~Gonzalez, D.~M.~Immig and A.~Weber acknowledge support by the HighRR research training group [GRK 2058].

Measurements leading to these results have been performed at the Test Beam 
Facility at DESY Hamburg (Germany), a member of the Helmholtz Association (HGF). We would like to thank the coordinators and support at DESY for the excellent test beam environment.

Further, we would like to thank the Paul Scherrer Institute for providing high rate test beams under excellent conditions.

\bibliographystyle{unsrt_collab_comma}
\bibliography{mybib}

\begin{thebibliography}{10}

\bibitem{Mu3e_TDR}
K.~Arndt et~al., {\em ``{Technical design of the phase I Mu3e experiment}''},
  arXiv:2009.11690, 2020.

\bibitem{Peric:2007zz}
I.~Peri\'c, {\em ``{A novel monolithic pixelated particle detector implemented
  in high-voltage CMOS technology}''}, Nucl.Instrum.Meth., \textbf{A582} 876,
  2007.

\bibitem{Augustin:2016hzx}
H.~Augustin et~al., {\em ``{The MuPix System-on-Chip for the Mu3e
  Experiment}''}, Nucl. Instrum. Meth., \textbf{A845} 194--198, 2017.

\bibitem{PIXEL:MuPix7}
H.~Augustin et~al., {\em ``{MuPix7-A fast monolithic HV-CMOS pixel chip for
  Mu3e}''}, Journal of Instrumentation, \textbf{11}(11) C11029, 2016.

\bibitem{Augustin:2016pwd}
H.~Augustin et~al., {\em ``{The MuPix Telescope: A Thin, high Rate Tracking
  Telescope}''}, JINST, \textbf{12}(01) C01087, 2017.

\bibitem{Augustin:2017guc}
H.~Augustin et~al., {\em ``{Irradiation study of a fully monolithic HV-CMOS
  pixel sensor design in AMS 180~nm}''}, Nucl. Instrum. Meth., \textbf{A905}
  53--60, 2018.

\bibitem{Augustin:2018ppf}
H.~Augustin et~al., {\em ``{Efficiency and timing performance of the MuPix7
  high-voltage monolithic active pixel sensor}''}, Nucl. Instrum. Meth.,
  \textbf{A902} 158--163, 2018.

\bibitem{Hiroshima19}
H.~Augustin et~al., {\em ``The MuPix sensor for the Mu3e experiment''}, Nucl.
  Instrum. Meth. A, \textbf{979} 164441, 2020.

\bibitem{LTU}
LTU, {LED Technologies of Ukraine -- {\tt http://ltu.ua/en/index/}}.

\bibitem{SpTAB}
M.~Oinonen et~al., {ALICE Silicon Strip Detector module assembly with
  single-point TAB interconnections}, In {\em {Proceedings, eleventh Workshop
  on Electronics for LHC and Future Experiments, Heidelberg, Germany, 12-16
  September 2005}}, pages 92--98, 2005.

\bibitem{Menzel2020}
M.~Menzel, {\em {Calibration of the MuPix10 pixel sensor for the Mu3e
  experiment}}, Bachelor thesis, Heidelberg University, 2020.

\bibitem{Hammerich2018}
J.~Hammerich, {\em {Analog Characterization and Time Resolution of a large
  scale HV-MAPS Prototype}}, Master thesis, Heidelberg University, 2018.

\bibitem{Pixel18}
H.~Augustin et~al., {\em ``{Performance of the large scale HV-CMOS pixel sensor
  MuPix8}''}, JINST, \textbf{14}(10) C10011, 2019.

\bibitem{Huth2018}
L.~Huth, {\em {A High Rate Testbeam Data Acquisition System and
  Characterization of High Voltage Monolithic Active Pixel Sensors}}, PhD
  thesis, Heidelberg University, 2018, https://www.psi.ch/mu3e/theses.

\end{thebibliography}

\end{document}